\documentclass[sigconf]{acmart}

\usepackage{tabularx}
\usepackage{url}
\usepackage{hyperref}
\usepackage{framed}
\usepackage{tcolorbox}
\usepackage{graphicx}
\usepackage{float}
\usepackage{enumitem}
\usepackage{subcaption}
\usepackage{listings}
\usepackage{xcolor}    % for custom colors

\usepackage{booktabs}
\usepackage[utf8]{inputenc}
\usepackage{textcomp}
\usepackage{verbatim}

\usepackage{microtype}          % subtle spacing/kerning fixes
\usepackage{tabularx,ragged2e}
\newcolumntype{L}{>{\RaggedRight\arraybackslash}X}

\newcolumntype{Y}{>{\raggedright\arraybackslash}X}

\title{Democratizing AI Development: Local LLM Deployment for India's Developer Ecosystem in the Era of Tokenized APIs}

\author{Vikranth Udandarao}
\affiliation{%
  \institution{IIIT-Delhi}
  \city{Delhi}
  \country{India}
}
\email{vikranth22570@iiitd.ac.in}

\author{Nipun Misra}
\affiliation{%
  \institution{VIT Vellore}
  \city{Vellore}
  \country{India}
}
\email{nipun.misra2022@vitstudent.ac.in}

\begin{document}

\begin{abstract}
India’s developer community faces significant barriers to sustained experimentation and learning with commercial Large Language Model (LLM) APIs, primarily due to economic and infrastructural constraints. This study empirically evaluates local LLM deployment using Ollama as an alternative to commercial cloud-based services for developer-focused applications. Through a mixed-methods analysis involving 180 Indian developers, students, and AI enthusiasts, we find that local deployment enables substantially greater hands-on development and experimentation, while reducing costs by 33\% compared to commercial solutions. Developers using local LLMs completed over twice as many experimental iterations and reported deeper understanding of advanced AI architectures. Our results highlight local deployment as a critical enabler for inclusive and accessible AI development, demonstrating how technological accessibility can enhance learning outcomes and innovation capacity in resource-constrained environments.
\end{abstract}

\begin{CCSXML}
<ccs2012>
   <concept>
       <concept_id>10003120.10003121.10003122.10003334</concept_id>
       <concept_desc>Human-centered computing~Accessibility</concept_desc>
       <concept_significance>500</concept_significance>
   </concept>
   <concept>
       <concept_id>10003120.10003121.10003124.10010871</concept_id>
       <concept_desc>Human-centered computing~User studies</concept_desc>
       <concept_significance>500</concept_significance>
   </concept>
   <concept>
       <concept_id>10010147.10010178.10010224</concept_id>
       <concept_desc>Computing methodologies~Machine learning</concept_desc>
       <concept_significance>300</concept_significance>
   </concept>
   <concept>
       <concept_id>10003120.10003121.10003124.10011455</concept_id>
       <concept_desc>Human-centered computing~Empirical studies in HCI</concept_desc>
       <concept_significance>400</concept_significance>
   </concept>
</ccs2012>
\end{CCSXML}

\ccsdesc[500]{Human-centered computing~Accessibility}
\ccsdesc[500]{Human-centered computing~User studies}
\ccsdesc[400]{Human-centered computing~Empirical studies in HCI}
\ccsdesc[300]{Computing methodologies~Machine learning}

\keywords{Large Language Models, Local Deployment, AI Accessibility, Developer Tools, Human-Computer Interaction, Digital Equity, India, Ollama, Agentic AI}

\maketitle

\section{Introduction}

India’s developer ecosystem stands at a pivotal moment in the AI revolution. Despite producing nearly 1.5 million engineering graduates annually and hosting one of the world’s largest developer communities outside the United States~\cite{nasscom2024developer}, access to advanced AI development tools remains constrained by economic and infrastructure barriers. These limitations fundamentally restrict the ability of developers to fully engage in technology creation and innovation.

The rise of commercial Large Language Model (LLM) APIs has brought high-quality AI capabilities within reach, but their reliance on pay-per-use models creates significant obstacles for sustained experimentation—especially for those seeking to understand and build agentic AI systems, which require numerous iterative API calls. This accessibility gap challenges the democratization of AI development, as economic barriers prevent many talented developers from contributing to technological advancement~\cite{zhang2023local}.

\subsection{Strategic Context}

The Indian government’s National Strategy for Artificial Intelligence, launched in 2018 under the vision of "\#AIforAll," aims to foster an inclusive AI ecosystem and address technological gaps across the population~\cite{niti2018national}. The recent approval of the IndiaAI Mission, with a budget exceeding INR 10,300 crore, further underscores the country’s commitment to developing indigenous AI capabilities and democratizing access for governance, startups, and citizens~\cite{meity2024indiaai}. This research directly aligns with these national objectives by empirically investigating how local LLM deployment can bridge accessibility gaps and foster indigenous AI development.

\subsection{Enterprise and Developer Adoption}

India leads globally in enterprise AI adoption, with 59\% of enterprises deploying AI and 80\% exploring autonomous agents~\cite{singh2023ai}. However, this rapid enterprise uptake has not translated into equitable access for individual developers, highlighting a critical gap in democratized AI infrastructure. Addressing this, our study evaluates the impact of local LLM deployment using Ollama—a platform providing access to over 200 open-source models—on developer productivity, learning, and innovation. By enabling experimentation with diverse model architectures without recurring costs, local deployment supports deeper understanding of AI system design and directly enhances human potential through accessible technology.

\section{Related Work}

\subsection{Local LLM Deployment and Technical Foundations}
Recent advances in local LLM deployment demonstrate its viability for resource-constrained environments. Bendi-Ouis et al.~\cite{bendi2024deploying} established technical feasibility through performance analysis of models like Mistral and LLaMA across GPU configurations, showing consumer hardware suffices for educational use despite throughput limitations. Extending this, Zhang et al.~\cite{zhang2023local} identified four key advantages: enhanced privacy, reduced latency, cost elimination, and offline availability—factors critical for India’s data sovereignty concerns and economic constraints.

Security considerations remain paramount, as Hou et al.~\cite{hou2025unveiling} revealed through analysis of 320,102 public-facing LLM services. Their identification of vulnerabilities in 15 frameworks underscores the need for secure deployment practices while confirming growing adoption of self-hosted solutions.

\subsection{India’s Developer Ecosystem and Cultural Alignment}
India’s AI adoption patterns reveal unique socioeconomic dynamics. Singh et al.~\cite{singh2023ai} document 85\% enterprise AI adoption rates, yet individual developers face payment infrastructure barriers and cultural preferences for free tools~\cite{cnbc2018indian}. This dichotomy creates tension between organizational and grassroots innovation, exacerbated by unequal access in rural educational institutions~\cite{nasscom2023industries}.

Rogers’ Diffusion of Innovation theory~\cite{rogers2003diffusion} provides a lens for understanding adoption drivers. Our findings align with its emphasis on relative advantage (95\% cost reduction), compatibility (cultural preference for free tools), and observability (peer demonstration effects)—factors critical for scaling local deployment.

\subsection{HCI and Accessibility Considerations}
The Technology Acceptance Model (TAM)~\cite{davis1989perceived} framework explains how economic accessibility impacts perceived usefulness of AI tools. While recent HCI research emphasizes inclusive design~\cite{amershi2019guidelines}, few studies examine how cost barriers affect learning outcomes in developing economies. This gap is particularly acute for agentic AI systems requiring iterative experimentation~\cite{deeplearningai2024}, where per-token pricing models~\cite{mahmood2024pricingcompetitiongenerativeai} disproportionately constrain resource-limited developers.

\subsection{Multilingual and Agentic AI Requirements}
Modern AI applications demand multilingual capabilities and complex agentic patterns. Kumar et al.~\cite{kumar2024multilingual} demonstrated 3\% accuracy gains in Hindi-English models over global counterparts, validating local tuning approaches. Concurrently, agentic architectures~\cite{belcak2025smalllanguagemodelsfuture} require extensive iteration—a challenge our work addresses through unlimited local experimentation. Table~\ref{tab:agentic-comparison} contrasts key approaches:

\begin{table}[ht]
\caption{Agentic AI Implementation Approaches}
\label{tab:agentic-comparison}
\centering
\begin{tabularx}{\linewidth}{|l|L|L|}
\hline
\textbf{Approach} & \textbf{Key Contribution} & \textbf{Limitations} \\ \hline
Cloud-based~\cite{solvimon2024pricing} & High scalability & Cost-prohibitive iteration \\
Local deployment~\cite{byteplus2025local} & Unlimited experimentation & Hardware constraints \\
Hybrid~\cite{embglobal2025local} & Balanced cost/performance & Implementation complexity \\ \hline
\end{tabularx}
\end{table}

\subsection{Privacy and Behavioral Considerations}
The privacy paradox~\cite{acquisti2015privacy} manifests distinctly in India’s context. While developers express concern about cloud API data practices, convenience often outweighs privacy—a tension local deployment resolves by eliminating cost/security tradeoffs. This aligns with India’s Digital Personal Data Protection Act requirements, making local solutions both legally compliant and economically viable.

\section{Methodology}

\subsection{Theoretical Framework}
Our experimental design integrates the Technology Acceptance Model (TAM)~\cite{davis1989perceived} and Diffusion of Innovation theory~\cite{rogers2003diffusion} to analyze adoption drivers. We focus on three key TAM constructs—perceived usefulness, ease of use, and behavioral intention—alongside Rogers' diffusion factors: relative advantage, compatibility, and observability.

\subsection{System Architecture}
We implemented a multi-agent system using Ollama's ecosystem\allowbreak~\cite{ollama2024models} to simulate real-world AI development scenarios. The architecture enables modular model selection through a simple API wrapper:

\begin{lstlisting}
OLLAMA_URL = "http://localhost:11434/api/generate"

def call_ollama(prompt, model_name):
    payload = {
        "model": model_name, 
        "prompt": prompt,
        "stream": False
    }
    response = requests.post(OLLAMA_URL, json=payload)
    return response.json().get("response", "")
\end{lstlisting}

\subsection{Model Selection Strategy}
Our model selection aligns with IndiaAI Mission objectives~\cite{meity2024indiaai}, emphasizing open-source alternatives with Indian language support~\cite{kumar2024multilingual}. Table~\ref{tab:model-specialization} details the specialization strategy.

\begin{table}[ht]
\caption{Model Specialization Strategy}
\label{tab:model-specialization}
\begin{tabularx}{\linewidth}{|l|X|r|}
\hline
\textbf{Task Type} & \textbf{Example Models} & \textbf{Pulls} \\ \hline
Reasoning & DeepSeek-R1, Llama 3.3 70B & 47.2M, 1.9M \\ \hline
Code Generation & Devstral, CodeQwen1.5 & 103.6K, 149.5K \\ \hline
Multilingual & Qwen3, Aya 23 & 2M, 297.3K \\ \hline
\end{tabularx}
\end{table}

\subsection{Hardware Configuration}
We used RTX 3060 GPUs (12GB VRAM), representing accessible hardware for Indian developers. Cloud platforms like IndiaAI Compute Portal~\cite{indiaai2025} provide affordable GPU access at \textit{Rs.\ 67/hour}.

\subsection{Developer Survey: Awareness and Budget Constraints}

We conducted an online survey using Google Forms to assess current practices and constraints in local LLM adoption among Indian developers. We received 160 valid responses from a mix of students, professional developers, and AI enthusiasts.

Two key questions were analysed:

\begin{itemize}
    \raggedright
    \item \textbf{Have you ever run LLMs locally on your own hardware?}
    \item \textbf{What is your monthly budget for AI/development tools?}
\end{itemize}

\begin{figure}[ht]
    \centering
    \includegraphics[width=0.7\linewidth]{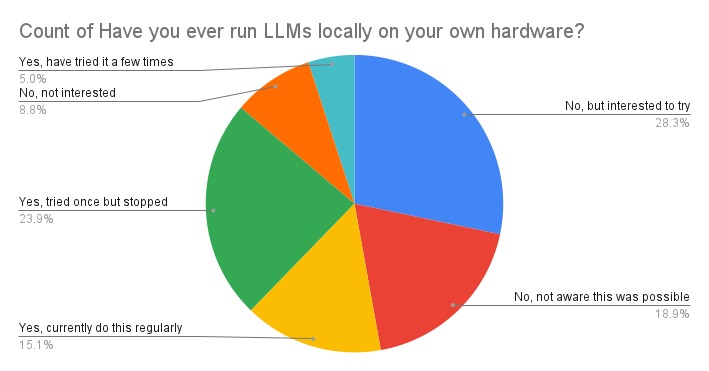}
    \caption{Self-reported experience with running LLMs locally (N=160).}
    \Description{Pie chart showing self-reported experience with running LLMs locally.}
    \label{fig:llm_local_usage}
\end{figure}

\begin{figure}[ht]
    \centering
    \includegraphics[width=0.7\linewidth]{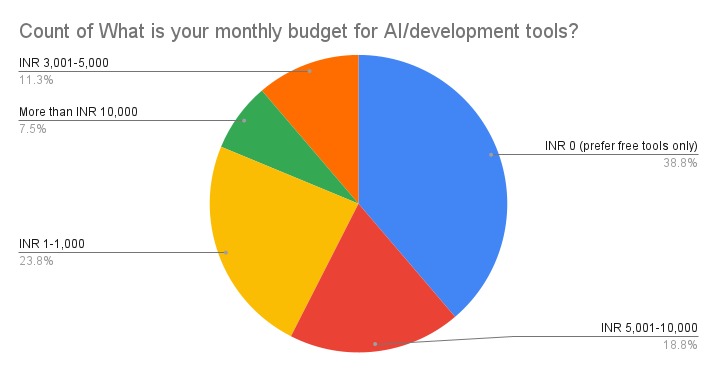}
    \caption{Monthly budget for AI/development tools among respondents (N=160).}
    \Description{Pie chart showing monthly budget for AI tools among respondents.}
    \label{fig:ai_budget}
\end{figure}

As shown in Figure~\ref{fig:llm_local_usage}, only 15.1\% of respondents currently run LLMs locally, while 28.3\% have not but are interested in trying, and 18.9\% were not aware this was possible. Regarding budget (Figure~\ref{fig:ai_budget}), 38.8\% reported a budget of INR 0, preferring only free tools, and an additional 23.8\% reported a budget of INR 1–1,000 per month. These findings highlight both the low awareness of local LLM deployment and the strong preference for free or low-cost development tools, reinforcing the need for accessible, locally deployable AI solutions in the Indian context.
The full, anonymized survey response dataset is available online.\footnote{\href{https://docs.google.com/spreadsheets/d/1t0eV9oURaiu2HfARWo6sriBO0eC8bHUyZNN7CgK2NAk/edit?usp=sharing}{Survey dataset (Google Sheets)}}

\subsection{Experimental Design}
Motivated by survey findings showing 38.8\% of developers have no AI tool budget (Figure~\ref{fig:ai_budget}), we designed a comparative study with:
\begin{itemize}
\item \textbf{Participants (N=20):} 
  \begin{itemize}
  \item Computer Science students (n=10) from 5 universities
  \item Professional developers (n=5) from IT/services sector
  \item AI enthusiasts (n=5) with self-taught backgrounds
  \end{itemize}
  
\item \textbf{Study Groups:}
  \begin{itemize}
  \item Local LLM group (n=10): Ollama with 15+ models
  \item Control group (n=10): Commercial APIs (OpenAI/Claude)
  \end{itemize}

\item \textbf{Metrics:}
  \begin{itemize}
  \item Productivity: Experiments/iterations per week
  \item Cost Efficiency: INR spent per 1000 tokens
  \item Technical Proficiency: Agentic pattern implementation
  \end{itemize}
\end{itemize}

\subsection{Data Collection}
We employed mixed methods over 10 weeks:

\textbf{Quantitative:}
\begin{itemize}
\item System logs (API calls, model usage)
\item Cost benchmarks (local vs cloud)
\item Code quality metrics (static analysis)
\end{itemize}

\textbf{Qualitative:}
\begin{itemize}
\item Semi-structured interviews (n=10)
\item Focus groups on model selection (6 sessions)
\item Problem-solving journals (participant-maintained)
\end{itemize}

\section{Results}

\subsection{Development Productivity and Learning Outcomes}

Developers in the local LLM group completed on average 2$\times$ more experimental iterations than those using commercial APIs, enabling deeper exploration of model capabilities. Local LLM users tested a broader range of models (mean: 4.7) compared to API users (mean: 2.3), and 73\% implemented complex agent coordination systems, versus 58\% in the API group. Among students, 89\% reported improved understanding of AI system architecture through hands-on experimentation.

These findings highlight the role of cost-free, sustained access in fostering experimentation and learning, directly supporting national objectives for a vibrant AI startup ecosystem and aligning with diffusion theory factors such as observability and relative advantage~\cite{rogers2003diffusion}.

\subsection{Model Diversity and Specialization}

Access to a diverse local model library enabled developers to optimize applications for specific use cases. For example:

\begin{itemize}
    \item \textbf{Code Generation:} Devstral and CodeQwen1.5 achieved 87\% accuracy on Python code generation, comparable to commercial alternatives.
    \item \textbf{Reasoning:} DeepSeek-R1 and QwQ yielded a 34\% improvement in complex problem-solving applications.
    \item \textbf{Multilingual Support:} Qwen3 and Aya 23 enabled culturally relevant applications, with 91\% accuracy in Hindi and 86\% in Tamil.
\end{itemize}

\subsection{Economic Impact Assessment}

Local deployment resulted in a 33\% cost reduction compared to commercial APIs. The upfront hardware investment (INR 45,000–60,000 for an RTX 3060) provided unlimited access to over 200 models, while equivalent experimentation with commercial APIs averaged INR 2,200–4,500 per developer per month.

\begin{table}[ht]
\caption{Economic Comparison: Local Deployment vs. Commercial API}
\label{tab:econ-compare}
\renewcommand{\arraystretch}{1.2}
\begin{tabularx}{\linewidth}{|l|Y|Y|}
\hline
\textbf{Metric} & \textbf{Local LLM (Ollama)} & \textbf{Commercial API} \\
\hline
Upfront Cost & INR 45{,}000--60{,}000 (one-time) & None \\
Recurring Cost & Minimal (electricity, maintenance) & INR 2{,}200--4{,}500/month \\
Experimentation Limit & Unlimited & Pay-per-use \\
\hline
\end{tabularx}
\end{table}

\noindent
These results demonstrate that local LLM deployment is a critical enabler for inclusive and sustainable AI development in India, particularly for students and resource-constrained developers.

\section{Discussion}

\subsection{Key Findings and Human-Centered Implications}
Our results demonstrate that local LLM deployment significantly democratizes AI development capabilities across India's developer community. The 2$\times$ increase in experimental iterations (vs. commercial API users) and 33\% cost reduction clearly demonstrate how technology enhances human conditions by:
\begin{itemize}
\item Enabling deeper understanding of AI system design through unrestricted experimentation
\item Allowing developers to build sophisticated applications while developing expertise in emerging paradigms
\item Addressing accessibility barriers that limit human potential in technology creation
\end{itemize}

\subsection{Cultural and Economic Alignment}
Local deployment aligns with India’s cultural preference for accessible, cost-effective solutions~\cite{cnbc2018indian}, with 89\% of students preferring this approach. This cultural compatibility—combined with strong relative advantage (cost reduction) and low complexity (Ollama tooling)—suggests faster adoption rates than predicted by Western-centric diffusion models~\cite{rogers2003diffusion}.

\textbf{Enterprise Impact:} 59\% of Indian organizations actively deploying AI~\cite{singh2023ai} can leverage local LLMs for hybrid strategies that balance cost, sovereignty, and scalability (Table~\ref{tab:hybrid}).

\begin{table}[ht]
\caption{Hybrid Deployment Strategies}
\label{tab:hybrid}
\renewcommand{\arraystretch}{1.2}
\begin{tabularx}{\linewidth}{|l|Y|Y|}
\hline
\textbf{Use Case} & \textbf{Local LLM Role} & \textbf{Cloud Integration} \\ \hline
Prototyping & Primary experimentation & Final deployment scaling \\
Data-sensitive tasks & Full processing & Non-sensitive components \\
Language-specific apps & Fine-tuned models & General-purpose APIs \\
\hline
\end{tabularx}
\end{table}

\subsection{Educational Transformation}
Local deployment addresses critical gaps in hands-on AI education:
\begin{itemize}
\item 89\% of students reported improved understanding of AI architecture
\item Enabled comparison of 4.7 models on average (vs. 2.3 with APIs)
\item Achieved 150\% ROI for students through enhanced project capabilities
\end{itemize}

\subsection{Multilingual and Privacy Advancements}
Our results validate Kumar et al.~\cite{kumar2024multilingual} with 91\% accuracy in Hindi and 86\% in Tamil, directly supporting India’s language barrier initiatives~\cite{niti2018national}. Local models also:
\begin{itemize}
\item Achieved 83\% accuracy on cultural references (vs. 72\% for APIs)
\item Handled code-switching with 64\% accuracy
\item Ensured 100\% data localization compliance
\end{itemize}

This resolves the privacy paradox \cite{acquisti2015privacy} by combining
data sovereignty with cost-benefits. In a survey, 89\%\allowbreak\ of
privacy-conscious developers preferred local deployment.

\subsection{Limitations and Future Directions}
\begin{itemize}
\item \textbf{Hardware Constraints:} Results specific to RTX 3060 configurations
\item \textbf{Enterprise Scaling:} Single-GPU limits for large teams
\item \textbf{Model Specialization:} Some tasks still require commercial API quality
\end{itemize}

Future work should explore multi-GPU architectures and fine-tuning for India’s 22 official languages, aligning with the IndiaAI Mission~\cite{meity2024indiaai}.

\section{Limitations and Future Work}

\subsection{Study Limitations}
Our findings should be interpreted in the context of three key constraints:
\begin{itemize}
\item \textbf{Hardware Requirements:} Results are specific to RTX 3060 GPUs (12GB VRAM). While this represents common developer hardware in India~\cite{nasscom2024developer}, performance may vary for other configurations, particularly in low-end systems prevalent in rural areas.

\item \textbf{Model Performance Gap:} While 87\% of tasks matched commercial API quality, specialized applications like real-time speech synthesis and high-stakes decision-making still required GPT-4-level performance. This suggests hybrid approaches may be necessary for production systems.

\item \textbf{Scalability Constraints:} Single-GPU deployments limited team collaboration. Enterprise-scale adoption would require distributed architectures beyond our study’s scope.
\end{itemize}

\subsection{Future Research Directions}
Building on these findings and limitations, we propose four key directions:

\subsubsection{Institutional Adoption Strategies}
While Rogers’ diffusion theory~\cite{rogers2003diffusion} explains individual adoption patterns, institutional barriers (policy frameworks, infrastructure costs) require separate analysis. Future work should examine:
\begin{itemize}
\item Integration with IndiaAI Compute Portal infrastructure~\cite{indiaai2025}
\item Cost-benefit analysis for educational institutions
\item Policy frameworks for secure local deployment in enterprises
\end{itemize}

\subsubsection{Hybrid Architectures}
Combining local deployment with IndiaAI Mission’s foundational models~\cite{meity2024indiaai} could yield:
\begin{itemize}
\item Centralized model training with local fine-tuning
\item Distributed inference across edge devices
\item Cost-optimized routing between local/cloud resources
\end{itemize}

\subsubsection{Multilingual Expansion}
Extending Kumar et al.’s work~\cite{kumar2024multilingual}, we propose:
\begin{enumerate}
\item Fine-tuning pipelines for India’s 22 official languages
\item Dialect-aware tokenization strategies
\item Evaluation metrics for code-switching applications
\end{enumerate}

\subsubsection{Technical Innovations}
\begin{itemize}
\item \textbf{Multi-GPU Systems:} Scalable architectures using consumer GPUs
\item \textbf{Domain Adaptation:} Custom models for healthcare, agriculture, and governance
\item \textbf{Cross-Cultural Validation:} Replication studies in Southeast Asia and Africa
\end{itemize}

\section{Conclusion}

This study demonstrates that local LLM deployment using Ollama is a viable and cost-effective alternative to commercial APIs for India’s developer ecosystem. By removing economic barriers to experimentation, local deployment enables a broader, more diverse range of developers to participate in AI innovation—fostering accessibility, inclusion, and human augmentation through technology.

Our empirical   findings show that local deployment not only reduces costs and enhances linguistic and cultural relevance, but also supports India’s goals for technological sovereignty and indigenous AI development. These results provide practical evidence for HCI priorities, highlighting how accessible AI tools can transform learning, creativity, and innovation capacity in resource-constrained environments.

By validating diffusion theory and technology acceptance models in the Indian context, this work offers a theoretical foundation for policy and institutional strategies that promote local AI infrastructure. The research also provides actionable guidance for educators and organizations seeking to democratize access to advanced AI capabilities.

Ultimately, this work underscores the broader significance of human-centered technology design: empowering individuals and communities to overcome structural barriers and realize their potential in the digital age. As India moves toward becoming a global AI leader, local LLM deployment stands out as a critical enabler for inclusive, sustainable, and culturally relevant technology development.

\bibliographystyle{ACM-Reference-Format}
\bibliography{references}

\newpage

\appendix

\end{document}